\documentclass[pra,superscriptaddress]{revtex4}
\usepackage{amsfonts}
\usepackage{amsmath}
\usepackage{amssymb}
\usepackage{graphicx}
\usepackage{epsfig}
\usepackage{citesort}

\begin{document}

\title{Measurement driven quantum evolution}

\author{L. Roa}
\affiliation{Center for Quantum Optics and Quantum Information, Departamento de F\'{\i}sica,
Universidad de Concepci\'{o}n, Casilla 160-C, Concepci\'{o}n,
Chile.}
\author{M. L. Ladr\'{o}n de Guevara}
\affiliation{Departamento de F\'{\i}sica, Universidad Cat\'{o}lica del Norte, Casilla 1280,
Antofagasta, Chile.}
\author{A. Delgado}
\affiliation{Center for Quantum Optics and Information, Departamento de F\'{\i}sica,
Universidad de Concepci\'{o}n, Casilla 160-C, Concepci\'{o}n,
Chile.}
\author{A. Klimov}
\affiliation{Departamento de F\'{\i}sica, Universidad de Guadalajara, Revoluci\'on 1500, 44420 Guadalajara, Jalisco,
Mexico.}

\date{\today}

\begin{abstract}
We study the problem of mapping an unknown mixed quantum state onto a known pure state without the use of unitary transformations. This is achieved with the help of sequential measurements of two non-commuting observables only. We show that the overall success probability is maximized in the case of measuring two observables whose eigenstates define mutually unbiased bases. We find that for this optimal case the success probability quickly converges to unity as the number of measurement processes increases and that it is almost independent of the initial state. In particular, we show that to guarantee a success probability close to one the number of consecutive measurements must be larger than the dimension of the Hilbert space. We connect these results to quantum copying, quantum deleting and entanglement generation.
\end{abstract}


\maketitle

\section{Introduction}

During the last two decades a major research effort has been conducted in the emerging field of quantum information theory \cite{Nielsen}. Much of this activity starts with the observation that the capacity of physical systems to process, store and transmit information depends on thir classical or quantum nature \cite{Landauer}. Quantum algorithms, that is, algorithms based on the laws of Quantum Mechanics, show an enhancement of information processing capabilities over their classical counterparts. A large collection of quantum communication protocols such as quantum teleportation \cite{Bennett}, entanglement swapping \cite{Zukowsky}, quantum cloning \cite{Wootters,Duan} and quantum erasing \cite{Pati} reveal new forms of transmitting and storing classical and quantum information. Most of these protocols have already been experimentally implemented \cite{Zhao1,Zhao2,Pan1,Pan2,Bouwmeester1,Boschi,Bouwmeester2}.

A common assumption concerning quantum algorithms and quantum communication protocols is the capacity of performing transformations belonging to a fixed but arbitrary set of unitary transformations together with measurements on a given basis. In this article we study the problem of mapping a mixed initial state onto a known pure state using measurements as the only allowed resource, that is, a measurement driven quantum evolution. We show how this problem connects naturally to generation of quantum copies, quantum deleting and entangled states generation. 

This article is organized as follows: in section \ref{2d} we study the problem considering states belonging to a two-dimensional Hilbert space. In section \ref{ddimension} we generalize to the case of a $d$-dimensional Hilbert space and show that mutually unbiased bases optimize the overall success probability. Section \ref{d-d and m/d} presents the case of $m$ target states in a $d$-dimensional Hilbert space. In section \ref{conclusions} we summarize our results.

\section{Two-dimensional case}   \label{2d}

Let us consider a quantum system described by a two-dimensional Hilbert space. Initially, the system 
is in a mixed state $\rho$. Our goal consists in mapping this state onto the known target
state $|\varphi\rangle$ by using quantum measurements as the only allowed resource. 

In order to accomplish this task we define a non-degenerate observable $\hat{\varphi}$. Its  spectral decomposition is
\begin{equation}
\hat{\varphi}=\lambda|\varphi\rangle\langle\varphi|
+
\lambda_{\perp}|\varphi_{\perp}\rangle\langle\varphi_{\perp}|,
\end{equation}
where the $|\varphi\rangle$ and $|\varphi_{\perp}\rangle$ states are eigenstates of $\hat{\varphi}$ with
eigenvalues $\lambda$ and $\lambda_{\perp}$ respectively. Thereby, the target state must belong to the spectral decomposition.

A measurement of the $\hat{\varphi}$ observable onto the $\rho$ state projects the system to the target
state $|\varphi\rangle$ with probability $p=\langle\varphi|\rho|\varphi\rangle$. In this case we succeed
and no further action is required. However, the process fails with probability $1-p$ when the measurement
projects the system onto the $|\varphi_{\perp}\rangle$ state. Since this state cannot be projected to $|\varphi\rangle$, the
target state, by means of another measurement of $\hat{\varphi}$, it is necessary to introduce a second
observable $\hat{\theta}$ whose nondegenerate eigenstates $|0\rangle$ and $|1\rangle$ are
\begin{eqnarray}
|0\rangle &=& \cos(\theta)|\varphi\rangle+e^{i\phi}\sin(\theta)|\varphi_{\perp}\rangle
\nonumber\\
|1\rangle &=& -e^{-i\phi}\sin(\theta)|\varphi\rangle+\cos(\theta)|\varphi_{\perp}\rangle,
\label{eq-1}
\end{eqnarray}
with $\theta$ and $\phi$ being real numbers.

\begin{figure}[t]
\begin{center}
\includegraphics[angle=360,width=0.40\textwidth]{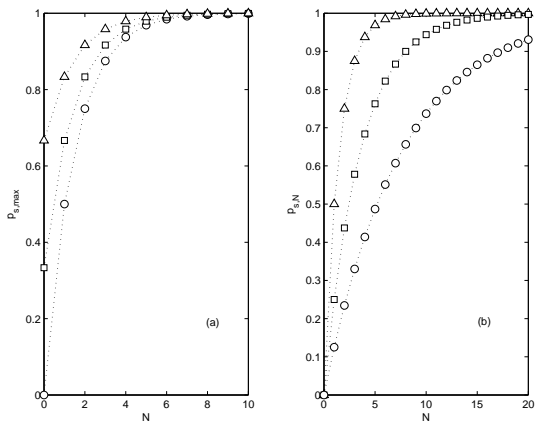}
\caption{Behavior of: (a) $p_{s,\max}$ as a function of N for three
values of $\langle\varphi|\rho|\varphi\rangle$: $2/3$(triangle), $1/3$ (square), $0$
(circle), (b) $p_{s,N}$ as a function of N for three values of $\theta$: $\pi/4$(triangle), $\pi/8$ (square), $\pi/12$ (circle), with
$\langle\varphi_\perp|\rho|\varphi_\perp\rangle=1$.}
\label{fig1}
\end{center}
\end{figure}

A measurement of $\hat{\theta}$ projects the $|\varphi_{\perp}\rangle$ state onto the
state $|0\rangle$ or  $|1\rangle$. Since both states have a component on the $|\varphi\rangle$
state, a second
measurement of $\hat{\varphi}$ allows us to project again, with a certain probability,
to the target state $|\varphi\rangle$. The probability $p^{\prime}$ that this procedure fails after
a first measurement of
$\hat{\varphi}$ but is successful after a consecutive measurement of the
$\hat{\theta}$ and $\hat{\varphi}$ operators is
\begin{equation}
\ p^{\prime}=\langle\varphi_{\perp}
|\rho|\varphi_{\perp}\rangle(\left\vert \langle0|\varphi_{\perp}
\rangle\right\vert ^{2}\left\vert \langle\varphi|0\rangle\right\vert ^{2}+
\left\vert \langle1|\varphi_{\perp}\rangle\right\vert ^{2}\left\vert
\langle\varphi|1\rangle\right\vert ^{2})= \frac{1}{2}\langle\varphi_{\perp}
|\rho|\varphi_{\perp}\rangle\sin^{2}(2\theta),
\end{equation}
where we have used Eq.(\ref{eq-1}). Then, the success probability in the sequence of measurements
$(M(\hat{\varphi})M(\hat{\theta}))M(\hat{\varphi})$ is
\begin{equation}
p+p^{\prime}=1-\langle
\varphi_{\perp}|\rho|\varphi_{\perp}\rangle (1-\frac{1}{2}\sin^{2}(2\theta)).
\end{equation}
Similarly, the success probability $p_{s}$ of
mapping the initial state $\rho$ onto $|\varphi\rangle$, the target state, after applying the consecutive measurement proceses $[M(\hat{\varphi})M(\hat{\theta})]^NM(\hat{\varphi)}$, that is, a measurement of
$\hat{\varphi}$ followed by $N$ measurement processes each one composed of $\hat{\theta}$ followed by $\hat{\varphi}$, is given by
\begin{equation}
p_{s,N}  = 1-\langle\varphi_{\perp}|\rho|\varphi_{\perp}\rangle\left(
1-\sum_{j=0}^{1}\left\vert \langle j|\varphi_{\perp}\rangle\right\vert
^{2}\left\vert \langle\varphi|j\rangle\right\vert ^{2}\right)^{N},
\end{equation}
or equivalently
\begin{equation}
p_{s,N}= 1-\langle\varphi_{\perp}|\rho|\varphi_{\perp}\rangle\left(  1-\frac{1}
{2}\sin^{2}(2\theta)\right)^{N}.
\label{pe}
\end{equation}
Clearly, the extreme values $\theta=0$ and $\theta=\pi/2$ correspond to observables $\hat{\varphi}$ and $\hat{\theta}$ defining the same basis. Consequently, in this case the success probability becomes simply $\langle\varphi|\rho|\varphi\rangle$. The expression (\ref{pe}) indicates that the success probability $p_{s,N}$ can be maximized by choosing $\theta=\pi/4$. In this case we obtain
\begin{equation}
p_{s,\max}=1-\frac{\langle\varphi_{\perp}|\rho|\varphi_{\perp}\rangle}{2^{N}
}.
\end{equation}
Fig. (\ref{fig1}a) illustrates the behavior of the maximum success probability $p_{s,\max}$ as a
function of $N$ for different values of $\langle\varphi|\rho|\varphi\rangle$. We observe
that $p_{s,\max}$ quickly converges to $1$ almost independently of the $\langle\varphi|\rho|\varphi\rangle$ even if the initial state $\rho$ belongs to a subspace orthogonal to $|\varphi\rangle$. For instance, in this particular case, if the success probability for the first measurement of $\hat{\varphi}$ vanishes, after four successive measurements of $\hat{\theta}$ and $\hat{\varphi}$ the success probability has increased to approximately $0.93750$, while after twelve measurement processes it reaches approximately the value $0.99976$. The fact that the success probability is maximized for $\theta=\pi/4$ indicates that in this case the $\hat{\theta}$ and $\hat{\varphi}$ observables define two mutually unbiased bases in a two-dimensional Hilbert space. Fig. (\ref{fig1}b) shows $p_{s,N}$ versus $N$ for $\theta$ equal to $\pi/12$ (circle), $\pi/8$ (square), and $\pi/4$ (triangle). Since mutually unbiased bases give the optimal process for each $N$, $p_{s,N}$ approaches to $1$ faster than in the other cases. As is apparent from Fig. (\ref{fig1}b), the convergence of the success probability strongly depends on the relation between the involved bases. In the following section we study this relation and generalize the results of this section to the $d$-dimensional case.

\section{$D$-dimensional case}  \label{ddimension}

Two noncommuting and nondegenerate observables defined on a $d$-dimensional Hilbert space can have at most $n$ equal eigenstates with $0\leq n\leq d-2$. Thus, on this $n$-dimensional subspace both observables can be well defined simultaneously, that is, the system can be described by one of the $n$ common eigenstates. However, on the $(d-n)$-dimensional subspace only one of them can be well defined. This property for two noncommuting observables turns up in a Hilbert space only when its dimension is higher that $2$. One can easily conclude that observables having some common eigendirections are not useful for our purpose. In this case the noncommutativity of the observables is a necessary but not sufficient condition, as in the two-dimensional case studied in the previous section. This motivates us to study the scheme of driving a quantum state by means of measurement in the general case of a $d$-dimensional Hilbert space.
       
We now generalize the previous results to the case of a target state, $|\varphi_1\rangle$, belonging to a
$d$-dimensional Hilbert space. Let $\{|\varphi_{1}\rangle,|\varphi_{2}\rangle,\dots,|\varphi_{d}\rangle\}$
and $\{|1\rangle,|2\rangle,\dots,|d\rangle\}$ be orthonormal bases defined by the spectral decompositions of the non-degenerate observables $\hat{\varphi}$ and $\hat{\theta}$ respectively. Initially the system is described by the $\rho$ state to be mapped onto the
known pure state $|\varphi_{1}\rangle$. The probability of success after $N$ measurement
processes of $\hat{\theta}$ followed by $\hat{\varphi}$, is given by
\begin{equation}
p_{s,N}=\langle\varphi_1|\rho|\varphi_1\rangle+
\sum_{i=2}^d\langle\varphi_i|\rho|\varphi_i\rangle
\sum_{k=1}^N\big[\prod_{n=1}^k \big(\sum_{j_1=2}^d \dots \sum_{j_n=2}^d p_{i,j_1}p_{j_1,j_2}
\dots p_{j_n,1}\big) \big],
\label{se}
\end{equation}
where $p_{k,j}$ is defined as
\begin{equation}
p_{k,j}=\sum_{i=1}^{d}\left\vert \langle i|\varphi_{k}\rangle\right\vert
^{2}\left\vert \langle\varphi_{j}|i\rangle\right\vert ^{2}.
\end{equation}
The process which maps the $|\varphi_{k}\rangle$ state with $k\neq1$ onto
the $|\varphi_{1}\rangle$ state is fundamental in this protocol because it is repeated
when we do not succeed. The success probability of this process is $p_{k,1}$.
This can be seen as an inner product between $A_i^k$ vectors whose $d$ real, non-negative
components are $|\langle i|\varphi_k\rangle|^2$ ($i=1,\dots,d$), that is, $p_{k,1}=\sum_{i=1}^d A_i^kA_i^1$.
This product is maximum when both vectors are parallel, which implies that $A_i^k=\alpha_k A_i^1$ for all $k$.
Since these vectors are real and the sum of their components is unitary, we deduce that
$\alpha_k=1$ for all $k$ and that $\left\vert \langle i|\varphi_{k}\rangle\right\vert
^{2}=\left\vert \langle i|\varphi_{k^{\prime}}\rangle\right\vert ^{2}$.
Therefore, we conclude that $\left\vert \langle i|\varphi_{k}\rangle\right\vert
=1/\sqrt{d}$ $\forall$ $i,k$. This property indicates that, in the optimal case, the two $\hat{\theta}$ and $\hat{\varphi}$ observables
define two mutually unbiased bases. An alternative proof can be obtained by noting that it suffices to optimize the first step. That is, we need to project a
$|\varphi_{\perp}\rangle$ state onto some element of the $\{|i\rangle\}$ basis in order to take the state out from the subspace orthogonal to the desired direction. The resulting density matrix of this process is
\begin{equation}
\rho=\sum_{i}|\langle i|\varphi_{\perp}\rangle|^{2}|i\rangle\langle i|.
\end{equation}
Now we look for the basis which leads to the state nearest to the target state $|\varphi_1\rangle$. This can be quantified by means of the Hilbert-Schmidt distance \cite{Lee}. In this case, we need to minimize the expression
\begin{equation}
D=\min||\rho-|\varphi_1\rangle\langle\varphi_1|||^{2}.
\end{equation}
Considering the $\rho$ state, this expression becomes
\begin{equation}
D=2(1-\sum_{i}|\langle i|\varphi_{\perp}\rangle|^{2}|\langle\varphi_1
|i\rangle|^{2}).
\end{equation}
Taking into account the properties of the above probabilities, it is clear
that the minimum distance is reached under the condition
\begin{equation}
|\langle i|\varphi_{\perp}\rangle|^{2}=|\langle\varphi_1|i\rangle|^{2},
\end{equation}
which means that the $\{|i\rangle\}$ basis must be complementary to the original $\{|\varphi_j\rangle\}$ basis. That is, the two required bases are related by means of the discrete Fourier transformation. In this scheme only two complementary bases are required, which can always be found \cite{Archer,Chaturvedi,Santhanam}. A different proof can be obtained by interpreting $p_{k,1}$ as a correlation function and considering
the property $p_{1}\geq p_{k}$ \cite{Reif}.

For mutually unbiased bases the success probability, Eq. (\ref{se}), simplifies
considerably to
\begin{equation}
p_{s,N}=1-\left(  1-\langle\varphi_{1}|\rho|\varphi_{1}\rangle\right)  \left(
1-\frac{1}{d}\right)^{N}.
\label{pk}
\end{equation}
In the limit, $d\gg 1$, this expression becomes
\begin{equation}
p_{s,N}=1-\left(1-\langle\varphi_{1}|\rho|\varphi_{1}\rangle\right)
e^{-\frac{N}{d}}.
\label{pke}
\end{equation}
Thus, in the case of higher dimensions, in order to reach a success probability
close to $1$, it is required that the number $N$ of measurement processes of the
$\hat{\theta}$ observable followed by $\hat{\varphi}$ must be larger than the dimension
$d$ of the Hilbert space. Otherwise, the term $\langle\varphi_{1}|\rho|\varphi_{1}\rangle$ entering in Eqs. (\ref{pk}) and (\ref{pke})
dominates.

We now proceed to obtain an average success probability which does not depend on the initial
pure state. This is achieved by integrating over the whole Hilbert space, that is
\begin{equation}
\overline{p}_{s,N}=\int d\psi p_{s,N},
\end{equation}
where $d\psi$ denotes the Haar integration measure and we consider initially pure states
only. In this case the average probability $\overline{p}_{s,N}$ is
\begin{equation}
\overline{p}_{s,N}=1-(1-\frac{1}{d})^N+(1-\frac{1}{d})^N\int d\psi |\langle \varphi_{1}|\psi\rangle |^2,
\end{equation}
where we have considered the case of mutually unbiased bases. The starting point is the identity \cite{Banaszek}
\begin{equation}
\int d\psi \langle\psi|n\rangle\langle k|\psi\rangle |\psi\rangle\langle\psi|=\frac{\delta_{n,k}\mathbb{I}+|n\rangle\langle k|}{d(d+1)},
\end{equation}
where $\{|n\rangle\}$ with $n=1,\dots,d$ is an arbitrary base for a $d$-dimensional
Hilbert space. Taking the trace of this identity we obtain
\begin{equation}
\int d\psi \langle\psi|n\rangle\langle k|\psi\rangle =\frac{\delta_{n,k}}{d}.
\end{equation}
Making $n=k$ and considering $|\varphi_{1}\rangle$ as belonging to the basis, we obtain, for any $|\varphi_{1}\rangle$ state, the identity
\begin{equation}
\int d\psi |\langle \varphi_{1}|\psi\rangle |^2=\frac{1}{d}.
\end{equation}
Thereby, the average success probability becomes
\begin{equation}
\overline{p}_{s,N}=1-(1-\frac{1}{d})^{N+1}.
\label{Average success probability}
\end{equation}
Thus, if we randomly select an initial state, the average success probability of mapping this state
onto the target state $|\varphi_1\rangle$ is given by $\overline{p}_{s,N}$.
In the limit of large $N$, $\overline{p}_{s,N}$ becomes
\begin{equation}
\overline{p}_{s,N}=1-e^{-\frac{N+1}{d}}.
\end{equation}
These results are equal to the case when the initial state $\rho$ is $I/d$, see Eqs. (\ref{pk}) and (\ref{pke}).

An interesting application of this result arises when we study the case of a target state $|\varphi_1\rangle$ belonging to a bipartite system, each system being described by a $d$-dimensional Hilbert space. In particular
\begin{equation}
|\varphi_1\rangle=\alpha|\psi\rangle|\psi_{\perp}\rangle+\beta|\psi_{\perp}\rangle|\psi\rangle.
\end{equation}
Assuming a factorized initial state of the form
\begin{equation}
\rho_i=\rho\otimes\rho,
\end{equation}
the success probability of the process which maps this state onto the $|\varphi_1\rangle$ state is given by
\begin{equation}
p_{s,N}=1-[1-\langle\psi|\rho|\psi\rangle(1-\langle\psi|\rho|\psi\rangle)(1+2\Re(\alpha\beta^*)|\gamma|^2)]
(1-\frac{1}{d^2})^N,
\end{equation}
where the $\gamma$ coefficient gives account of the initial decoherence process affecting the $\rho$ state. This coefficient relates the diagonal coefficients of $\rho$ to the non-diagonal coefficients through the relation
\begin{equation}
\langle\psi|\rho|\psi_{\perp}\rangle=\gamma\sqrt{\langle\psi|\rho|\psi\rangle \langle\psi_{\perp}|\rho|\psi_{\perp}\rangle},
\end{equation}
with $0\leq|\gamma|\leq 1$.
It can be shown that the success probability can be upper bounded as
\begin{equation}
p_{s,N}\leq1-[1-\frac{1}{4}(1+2\Re(\alpha\beta^*)|\gamma|^2)]
(1-\frac{1}{d^2})^N.
\end{equation}
Thereby, when $\Re(\alpha\beta^*)>0$, the maximum probability for fixed $N$ is achieved under the
condition $|\gamma|^2=1$, that is, for a pure initial state. However, if $\Re(\alpha\beta^*)<0$, the
probability is maximum when  $|\gamma|^2=0$. This means that, for states fulfilling the
condition $\Re(\alpha\beta^*)<0$, such as the singlet state, the success probability is higher in
the case of total initial decoherence than in any other case, corresponding the smallest probability
to an initially pure state.

This scheme can also be connected to the application of quantum erasure. If we fix the target state, and consequently the $\hat{\theta}$ and $\hat{\varphi}$ operators,  then the sequence of measurements will map any initial state onto that same target state. Thereby, the overall effect will correspond to probabilistically erasing the information content of the initial state. The success probability of this probabilistic erasure will be given by Eq. (\ref{Average success probability}).  

\section{Generalization to $m$ orthogonal target states} \label{d-d and m/d}

The above results can be generalized to the case of $m$ orthogonal target states belonging
to a $d$-dimensional Hilbert space. Here we consider again the
bases $\{|\varphi_{1}\rangle,|\varphi_{2}\rangle,\dots,|\varphi_{d}\rangle\}$ and $\{|1\rangle,|2\rangle,
\dots,|d\rangle\}$ of the observables $\hat{\varphi}$ and $\hat{\theta}$ respectively. Our aim is to map
the initial state $\rho$ onto any of the target states $\{|\varphi_{1}\rangle,\dots,|\varphi_{m}\rangle\}$.
After measuring the $\hat\varphi$ observable and failing, the state of the system is in one of
the $\{|\varphi_{m+1}\rangle,\dots,|\varphi_{d}\rangle\}$ states. The probability of mapping the system
from any one of these states onto any one of the target states by a consecutive measurement of
the $\hat{\theta}$ and $\hat{\varphi}$ observables is given by
\begin{equation}
p=\sum_{j=m+1}^d\sum_{k=1}^m
p_{j,k}.
\end{equation}
This probability can be written as a sum of $(d-m)m$ scalar products of the $A_i^j$ vectors defined
in the previous section, that is
\begin{equation}
p=\sum_{j=m+1}^d\sum_{k=1}^m\sum_{i=1}^d A_i^j A_i^k.
\end{equation}
The maximum value of this quantity is achieved when each scalar product involves two parallel vectors, that is
\begin{equation}
A_i^j=\alpha_{j,k} A_i^k\qquad \forall j=m+1,\dots,d \:{\rm and}\: k=1,\dots,m \:{\rm and}\: i=1,\dots,d.
\end{equation}
Since the sum of the elements of each $A_i^j$ vector is unity, we obtain $\alpha_{j,k}=1$ $\forall j,k$. Thus,
it holds that $A_i^j=A_i^k$, that is, all the $A_i^j$ vectors are equal. This implies that any state $|i\rangle$ has
the same projection onto all the states belonging to the $\{|\varphi_{1}\rangle,|\varphi_{2}
\rangle,\dots,|\varphi_{d}\rangle\}$ basis, which is possible only if $|\langle\varphi_j|i\rangle|^2=1/d$.
Therefore, the $\hat{\varphi}$ and $\hat{\theta}$ observables define mutually unbiased bases. Considering
these types of bases, which optimize the process, the probability of mapping the initial state $\rho$ onto any of $m$ states
$|\varphi_1\rangle,\dots,|\varphi_m\rangle$ after $N$ measurement
processes of $\hat{\theta}$ and $\hat{\varphi}$ is given by
\begin{equation}
p_{s,N}=1-\left(1-\sum_{k=1}^m\langle\varphi_k|\rho|\varphi_k\rangle\right)\left(1-\frac{m}{d}\right)^N.
\end{equation}

As an application, let us suppose that we want to generate $l$ copies of
each of the $m$ orthogonal states $|\psi_1\rangle,|\psi_2\rangle,\dots,|\psi_m\rangle$, belonging
to a $d$-dimensional Hilbert space. So, we can assume that
the $m$ target states belong to a multipartite system composed of $l$ identical systems
and that they have the form
$|\varphi_i\rangle=|\psi_i\rangle_1\otimes|\psi_i\rangle_2 \otimes\dots\otimes|\psi_i\rangle_l$, where
$i=1,\dots,m$.
Then the probability of generating any of these ``state-copies''
after $N$ processes of measurement of $\hat{\theta}$ followed by $\hat{\varphi}$ is
\begin{equation}
p_{s,N}=1-\left(1-\sum_{k=1}^m \langle\psi_k|\rho|\psi_k\rangle^l\right)\left(1-\frac{m}{d^l}\right)^N,
\end{equation}
where we have assumed that the initial state is factorized and that each of the $l$ systems is in the $\rho$
state. Independently of the initial condition, this probability is closer to unity when $d^l\gg m$ and $Nm\gg d^l$.
On the other hand, randomly selecting an initially pure state the average success probability of mapping
this state onto one of the  $m$ target ``state-copies'' is given by the expression
\begin{equation}
\overline{p}_{s,N}=1-\left(1-\frac{m}{d^l}\right)^{N+1},
\end{equation}
which in the limit $d^l\gg m$ behaves as
\begin{equation}
\overline{p}_{s,N}=1-\exp\left(\frac{m(N+1)}{d^l}\right).
\end{equation}
Thus, in this limit the more the number $l$ of copies, the probability of success decreases or converges more slowly to $1$.

\section{Conclusions} 
\label{conclusions}
We have studied a scheme to map an unknown mixed state of a quantum system onto an arbitrary state belonging to a set of known pure quantum states. This scheme is based on a sequence of measurements of two noncommuting observables. The target states are eigenstates of one of the two observables, while the other observable maps the states out of the subspace orthogonal to the one defined by the target states. The success probability turns out to be maximal under the condition that the observables define mutually complementary bases. In other words, both required bases are always related by a discrete Fourier transform. We have also shown that these results hold in the case of arbitrary but finite dimensions. The scheme consists of applying a measurement of the $\hat{\varphi}$ observable followed by $N$ measurement processes each one composed of $\hat{\theta}$ followed by $\hat{\varphi}$, this is, $(\hat{\varphi}\hat{\theta})^N\hat{\varphi}$. The target states belong to the spectral descomposition of the $\hat{\varphi}$ observable.
The success probability quickly converges to unity when the number $N$ of the sequences of measurement processes is larger than the dimension of the Hilbert space. We have connected these results to the generation of quantum copies, quantum deleting and pure entangled states generation. The extension of these results to the case of continuous variables is under study.
 
\acknowledgements
This work was supported by Grants FONDECYT No. 1030671, FONDECYT No. 1040385, FONDECYT No. 1040591, and Milenio ICM P02-49F. A. D. thanks Fundaci\'on Andes.

\end{document}